\documentstyle[prl,aps]{revtex}

\begin{document}

\title{Anomalous Acceleration Effects for Neutrons}

\author{Brett D. Altschul\footnote{baltschu@mit.edu}}
\address{Department of Mathematics \\
Massachusetts Institute of Technology \\
Cambridge, Massachusetts 02139}

\date{\today}

\maketitle

\begin{abstract}
Comparing the Dirac Hamiltonians for a neutron subjected to either a
Schwartzchild gravitational field or a uniform acceleration, we observe that
the difference between the two is precisely the sort that might be eliminated
by the introduction of a new quantum number. The origin of this quantum number
lies in the noncommutation of an acceleration with the quark operators that
constitute the neutron. We show that the term containing the new quantum number
only acts on very long length scales. Furthermore, the symmetries of an
acceleration prevent the effects of this term from being periodic.
\end{abstract}

\narrowtext
\twocolumn

\section{Introduction}

Any laboratory on the Earth simultaneously experiences both acceleration and
gravity. Einstein's equivalence principle states that, neglecting curvature
effects, the acceleration and gravitation are indistinguishable. The success of
General Relativity is a testament to the accuracy of this statement in the
classical regime. The equivalence principle has also been tested in quantum
mechanical phenomena. First, Colella et al.\@~\cite{ref-colella} performed
their celebrated neutron interferometry experiment, observing a
gravitationally-induced phase shift. Using similar apparatus, Bonse and
Wroblewski~\cite{ref-bonse} found the same phase shift when their
interferometer was uniformly accelerated. Together, these experiments provided
some confirmation of the equivalence principle in quantum mechanics.

These two neutron interferometry experiments did not measure any spin-dependent
effects, leaving open the question of whether spin may be affected differently
by acceleration and uniform gravitation. Varj\'{u} and Ryder~\cite{ref-varju}
have suggested a particular spin effect that may be able to distinguish between
the two. They compared two Dirac (spin 1/2) Hamiltonians. One was the
Hamiltonian for a particle in a Schwartzchild gravitational field, drawing on
earlier calculations by Fischbach, et al.\@~\cite{ref-fischbach}. The other
Hamiltonian represented a Dirac particle subjected to a uniform acceleration,
as found by M.~S.~Altschul~\cite{ref-maltschul} and later by Hehl and
Ni~\cite{ref-hehl}. An apparent difference between these Hamiltonians calls
into question whether the equivalence principle holds for spin effects.

We will show that that an additional accelerational effect could eliminate the
difference between the two Hamiltonians, restoring the equivalence principle.
However, it is natural that this effect has not yet been observed, since it
acts on very long length scales.

\section{Comparison of Hamiltonians}

In comparing uniform gravitation and acceleration, we will evaluate the
Hamiltonians to first order. For the acceleration, that means to first order in
${\bf a}$, the acceleration vector. For the gravitational Hamiltonian, we will
work to first order in ${\bf g}$, where $g_{i}=-\frac{\partial \Phi}{\partial
x^{i}}c^{2}$. Here, $\Phi=\frac{GM}{c^{2}r}$, so ${\bf g}=-\frac{\Phi c^{2}}
{r^{2}}{\bf x}$ is just the normal gravity vector. Each of these Hamiltonians
may be found by using three successive Foldy-Wouthuysen
transformations~\cite{ref-foldy}.

The gravitational Hamiltonian found by Varj\'{u} and Ryder takes the form
\begin{eqnarray}
H&=&\beta mc^{2}-\beta m{\bf g}\cdot {\bf x}+\frac{\beta}{2m}p^{2}-\frac{\beta}
{2mc^{2}}{\bf p}\cdot({\bf g}\cdot {\bf x}){\bf p}\nonumber \\
& &+\frac{\hbar\beta}{2mc^{2}}\mbox{{\boldmath $\sigma$}}\cdot({\bf g}\times
{\bf p})-\frac{\beta}{mc^{2}}({\bf p}\cdot {\bf g})({\bf x}\cdot {\bf p}).
\label{eq-ghamil}
\end{eqnarray}
The accelerational Hamiltonian has been found in two different (and both
flawed) ways. The result was
\begin{eqnarray}
H&=&\beta mc^{2}+\beta m{\bf a}\cdot {\bf x}+\frac{\beta}{2m}p^{2}+\frac{\beta}
{2mc^{2}}{\bf p}\cdot({\bf a}\cdot {\bf x}){\bf p}\nonumber \\
& &+\frac{\hbar\beta}{4mc^{2}}\mbox{{\boldmath $\sigma$}}\cdot({\bf a}\times
{\bf p}).
\label{eq-ahamil}
\end{eqnarray}

The equivalence principle indicates that these two Hamiltonians should be
identical when we set ${\bf a}=-{\bf g}$, except for tidal terms in
(\ref{eq-ghamil}). On initial inspection, this does not appear to be the case.
There are two differences. The last term of (\ref{eq-ghamil}) is entirely
absent from (\ref{eq-ahamil}). Although only first order in ${\bf g}$, this
term is actually a tidal term, as Varj\'{u} and Ryder demonstrated, so its
absence from (\ref{eq-ahamil}) is expected.

The other difference between the Hamiltonians less easily dismissed. While
(\ref{eq-ghamil}) includes the term
\begin{equation}
\label{eq-gterm}
\frac{\beta\hbar}{2mc^{2}} \mbox{{\boldmath $\sigma$}}\cdot({\bf g}\times
{\bf p}),
\end{equation}
(\ref{eq-ahamil}) contains
\begin{equation}
\label{eq-aterm}
\frac{\beta\hbar}{4mc^{2}} \mbox{{\boldmath $\sigma$}}\cdot({\bf a}\times
{\bf p}).
\end{equation}
When we equate ${\bf a}=-{\bf g}$, these two terms differ by a factor of $-2$.
It appears that this difference could allow an observer to distinguish between
the acceleration and the gravitation. However, our closer analysis indicates
that this may not be the case.

\section{Anomalous Acceleration Moment}

The Hamiltonian (\ref{eq-ahamil}) was first derived by purely electromagnetic
means, for a charged particle in a uniform electric field. Then
(\ref{eq-aterm}) enters as a charge-dependent term, vanishing for a neutral
particle. The behavior of (\ref{eq-aterm}) in this regard is the same as that
of the spin magnetic moment term in a uniform magnetic field ${\bf B}$.
However, we know that the spin magnetic moment is actually an additional
parameter of the Dirac Hamiltonian; it is not zero for neutral particles with
internal structure. Since ${\bf a}$ and ${\bf B}$ have the same
three-dimensional vector group structure, there should be an analogous
parameter in the accelerational Hamiltonian.

In fact, when we looks at the infinitesimal generator for an acceleration, we
may identify this new parameter. As shown in~\cite{ref-maltschul}, the
generator has the form
\begin{equation}
\label{eq-chisimple}
\mbox{{\boldmath $\chi$}}=\mbox{{\boldmath $\chi$}}_{\Gamma^{0}}+
\frac{\beta\hbar}{4c^{2}}\,\mbox{\boldmath $\sigma$}\times\mbox
{{\boldmath $\nabla$}}_{p}+f_{\chi}.
\end{equation}
For small values of ${\bf a}$ this generates the Hamiltonian (\ref{eq-ahamil})
from the free particle Hamiltonian $H_{0}$ by $H=H_{0}+[{\bf a}\cdot
\mbox{{\boldmath $\chi$}},H_{0}]$. Under this prescription, the first term of
{\boldmath $\chi$} generates the ``classical'' potential energy $\beta m
{\bf a\cdot x}$, and $f_{\chi}$ commutes with $H_{0}$. The second term,
containing {\boldmath $\sigma$}, is the one of interest. We will call this term
$\mbox{{\boldmath $\chi$}}'$. It is $\mbox{{\boldmath $\chi$}}'$ that generates
the problematic term (\ref{eq-aterm}).

From the form of (\ref{eq-chisimple}), it is clear that there is a free
parameter. The coefficient of {\boldmath $\chi'$} may be freely adjusted
without affecting any lower-order terms in the Hamiltonian. An ``anomalous
acceleration moment,'' $\mu_{a}$ may be introduced. The presence of $\mu_{a}$
changes the generator according to
\begin{equation}
\mbox{{\boldmath $\chi$}}'=(1+\mu_{a})\frac{\beta\hbar}{4c^{2}}\,
\mbox{\boldmath $\sigma$}\times\mbox{{\boldmath $\nabla$}}_{p}.
\end{equation}
The variation of $\mu_{a}$ creates a one-parameter family of
representations of the acceleration group.

It remains to consider for what particles this anomalous contribution may be
nonzero and significant. For charged particles, the contribution is certainly
not significant. Attempts to accelerate such particles uniformly are prevented
by the emission of electromagnetic radiation and the induced radiative reaction
force~\cite{ref-jackson}. For neutral particles, this problem does not appear,
so we shall consider neutrons.

From the pure Dirac standpoint, the neutron should have no magnetic moment.
However, there is a neutron magnetic moment caused by QCD. Similarly, there
may be an nonzero anomalous acceleration moment for the neutron. In fact, it is
expected to be nonzero, for the following reason. The value  $\mu_{a}=0$ is
required by the conditions of Lorentz invariance and charge
conservation~\cite{ref-jaffe}. However, these conditions do not hold under
acceleration. Since a charge density causes acceleration, the charge and
acceleration operators do not commute. More specifically, the quark operators
$\psi_{q}^{\dag}$ that compose a neutron do not commute with acceleration, so
the quark content of an accelerated neutron is not well-defined. We may
estimate the effective commutator from the fact that a free quark would induce
a force $F\approx 10^{22}$~GeV/cm~\cite{ref-rolnick} on every
differently-colored quark in the universe. This yields an order of magnitude of
\begin{equation}
|[a,\psi_{q}^{\dag}]| \sim\frac{2}{3}\frac{F}{m_{q}}\approx 10^{36}\,
{\rm cm\cdot s^{-2}},
\end{equation}
since the quark mass $m_{q}$ is a few MeV\,$\cdot\,c^{-2}$ for the $u$ and
$d$ quarks. The lowest-order opportunity for terms stemming from this
commutator to enter the Hamiltonian is through $\mu_{a}$.

This noncommutation is what interferes with Hehl and Ni's derivation of the
accelerational Hamiltonian, in which they simply replace the partial
derivatives in the Dirac equation with covariant derivatives. Although
Varj\'{u} and Ryder use this same substitution to find the gravitational
Hamiltonian, the noncommutation of ${\bf g}$ with the quark operators lies in
the realm of quantum gravity, and so should be negligible. So the existence of
$\mu_{a}$ allows for the restoration of the equivalence principle.  In fact,
the equivalence principle provides a specific prediction of its value,
$\mu_{a}=-3$.

\section{Length Scale for the Acceleration Moment}

It is important to investigate the dynamical effects caused by $\mu_{a}$. We
shall begin this process by examining the length scale on which
$\mbox{{\boldmath $\chi$}}'$ operates. The order of magnitude of $\mu_{a}$
will probably not deviate too much from unity, since it is a dimensionless
parameter of nuclear structure. This is certainly true if $\mu_{a}$ takes the
value indicated by the equivalence principle.

To get a length scale from $\mbox{{\boldmath $\chi$}}'$, we must insert the
quantum-mechanical prescription that $i\hbar\mbox{{\boldmath $\nabla$}}_{p}=
{\bf x}$, to get
\begin{equation}
\label{eq-chiwithx}
\mbox{{\boldmath $\chi$}}'=(1+\mu_{a})\frac{\beta}{4ic^{2}}\,
\mbox{\boldmath $\sigma$}\times{\bf x}.
\end{equation}
The $i$ disappears when we take $\exp(i{\bf a}\cdot\mbox{{\boldmath $\chi$}})
\approx 1+i{\bf a}\cdot\mbox{{\boldmath $\chi$}}$. The terms multiplying
${\bf x}$ in $\mbox{{\boldmath $\chi$}}'$ give us a length scale for the
action of this effect in the Hamiltonian. Although the generator is only
correct for infinitesimal accelerations, it should give the right dependence
for the length scale. We drop $\beta$ and {\boldmath $\sigma$}, since they only
take the values $\pm 1$. This leaves
\begin{displaymath}
\frac{|1+\mu_{a}|}{4c^{2}}\,{\bf x}.
\end{displaymath}
Since the original dimensionless quantity for generating the Hamiltonian was
${\bf a}\cdot\mbox{{\boldmath $\chi$}}$, the length scale must depend on
$a=|{\bf a}|$. The final expression for determining the length scale is then
\begin{equation}
{\bf a}\cdot\mbox{{\boldmath $\chi$}}'\sim\frac{|1+\mu_{a}|}{4c^{2}}\,{\bf
|x||a|}.
\end{equation}
Therefore, the characteristic length scale for the accelerated neutron is
\begin{equation}
x_{a}=\frac{4c^{2}}{|1+\mu_{a}|a}.
\end{equation}
In cgs units, this has the numerical value of
\begin{equation}
\label{eq-scale}
x_{a}=\frac{3.6\times 10^{21}\, {\rm cm}}{|1+\mu_{a}|a}.
\end{equation}

For reasonable values of $a$ and $|1+\mu_{a}|$, the scale of $x_{a}$is
astrophysical. A natural place to look for this effect would be the solar wind,
since the solar wind contains many neutrons and it undergoes continuous
acceleration. The gravitational acceleration at the surface of the sun is
$a_{\odot}=GM_{\odot}/R_{\odot}^{2}=2.7\times 10^{4}$~cm\,$\cdot\,$s$^{-2}$. If
$|1+\mu_{a}|=2$, this indicates a length of $x_{a}=6.6\times 10^{16}$~cm or
0.036~LY. This is a substantial distance. Moreover, this is not the flight
distance of the neutrons from the sun. Instead, it represents their retardation
by the sun's gravitational attraction. The actual flight distance would be even
larger, and the sun's gravitational acceleration is certainly not uniform over
the distance. Thus, it is almost certainly not possible to observe any effect
on the solar wind neutrons.

Since the length scale due to the sun's gravitational acceleration is too
large, it is natural to consider situations in which $a$ is substantially
greater.  In particular, we will look at the acceleration at the surface of
a neutron star, with mass $1.5M_{\odot}$ and radius $10^{6}$ cm. For this body,
$a_{N\- S}=2.0\times 10^{14}$~cm\,$\cdot$\,s$^{-2}$. In this case, the value of
$x_{a}$ is $9.1\times 10^{6}$~cm, substantially smaller.  However, this is
still too large for the gravitational field to be nearly uniform. In any case,
it is unlikely that particles at a neutron star's surface would be free from
far stronger interactions, so the effect is almost certainly not significant
in this case, either.

These calculated length scales indicate that the term of the Hamiltonian
containing $\mu_{a}$ does not have a sizeable effect except at very long
distances or very high resolutions. Thus, it is natural that $\mu_{a}$ has not
yet been observed in any experiment. Moreover, this term of the Hamiltonian
probably has no direct astrophysical implications.

\section{Acceleration Moment Effects Allowed by Symmetry}
\label{sec-symm}
At this point, it is not at all clear what sort of effects will occur on the
length scale $x_{a}$. Fischbach et al.\ identify the term (\ref{eq-gterm}) as
a standard spin-orbit coupling term, since
\begin{equation}
\mbox{{\boldmath $\sigma$}}\cdot({\bf g}\times{\bf p})=-\frac{\Phi c^{2}}
{r^{2}}\,\mbox{{\boldmath $\sigma$}}\cdot({\bf x}\times{\bf p})=-\frac{2\Phi
c^{2}}{\hbar r^{2}}\,{\bf L\cdot S}.
\end{equation}
This is correct in general, but it is not useful in the approximation of a
uniform gravitational field ${\bf a}=-{\bf g}$, since the meaningful behavior
of ${\bf L= x\times p}$ relies on ${\bf x}$ not remaining constant.  We shall
therefore work only with (\ref{eq-aterm}), which is clearly not a simple
${\bf L\cdot S}$ interaction. In fact, we shall find that the simple precession
characteristic of ${\bf L\cdot S}$ interactions does not occur in this problem.
Our arguments will be qualitative, based on the symmetries of the situation, so
the numerical difference between (\ref{eq-aterm}) and (\ref{eq-gterm}) is not
significant.

As noted in~\cite{ref-maltschul}, $\mbox{{\boldmath $\chi$}}'$ does not affect
the space-time trajectory of a neutron. However, $\mbox{{\boldmath $\chi$}}'$
may affect the spin vector, similar to the effect of a magnetic field,
${\bf B}$. If we express the direction of the spin vector as $(\theta,\phi)$
and apply the magnetic field ${\bf B}$ along the $z$-axis, $\phi$ precesses.
However, the acceleration vector ${\bf a}$ possess a higher symmetry than
${\bf B}$, making the situation more complex.

The symmetry point group of ${\bf B}$ is $SO(2)$, since the physics are
unchanged by rotations about the $z$-axis. The acceleration ${\bf a}$ exhibits
the same properties. However, the magnetic field is an axial vector, changing
sign under reflection. Acceleration is a vector, so its symmetry point group is
$O(2$).

We will consider the action of a known acceleration ${\bf a}$, suitably
parameterized. The operation of the acceleration on the spin vector 
$(\theta,\phi)$ must commute with the elements of $O(2)$. This means that
$\phi$ must remain constant, since no rotation about the $z$-axis commutes
with the reflections in $O(2)$. So the only rotations allowed are in the polar
angle $\theta$. These rotations may not take $\theta$ to 0 or $\pi$, since
$\phi$ is indeterminate at these poles, so the transformation would not be
invertible. If $\theta$ is the only coordinate changing, it can not oscillate,
since it is restricted to the interval $(0,\pi)$, and the transformation
induces a rotation of definite direction at each point of this interval.

Changes in $\theta$ must rotate the spin vector either toward or away from the
$z$-axis, the direction of acceleration. Although the rate of this rotation may
vary, it must satisfy
\begin{equation}
\dot{\theta}|_{\theta=\theta_{0}}=\dot{\theta}|_{\theta=\pi-\theta_{0}}
\label{eq-theta}
\end{equation}
in order to be consistent with the action of the inverse transformation. The
inverse, which would correspond to acceleration along the $-z$-axis, inverts
$\dot{\theta}$. Symmetry dictates that if, at the angle $\theta_{0}$, the
regular rotation was toward the $z$-axis, the inverse rotation at the angle
$\pi-\theta_{0}$ must be toward the $-z$-axis at the same rate. Since the
inverse transformation just reverses the rotation direction, the regular
rotation at $\pi-\theta_{0}$ is the same as it is at $\theta_{0}$.

Although we have not demonstrated this fact, it seems likely that the range of
rotation of $\theta$ is the entire range $(0,\pi)$, with the sign of $\dot
{\theta}$ the same everywhere. This would result in the spin vector
continuously rotating towards or away from the direction of the acceleration,
with the angular frequency symmetric about $\theta=\frac{\pi}{2}$ and gradually
decreasing as $\theta$ approaches the pole.

\section{Conclusions}

The existence of $\mu_{a}$ does not ensure the validity of the equivalence
principle. $\mbox{{\boldmath $\chi$}}'$ itself gives us no indication of the
magnitude of the ``anomalous acceleration moment.'' However, $\mu_{a}$ does
provide a convenient method for restoring the equivalence principle, if
$\mu_{a}=-3$. This specific prediction gives us another avenue for analysis of
both $\mu_{a}$ and the equivalence principle. The particular value of $\mu_{a}$
is a property of the quark operators under acceleration. By examining the
problem from the QCD standpoint, it may be possible to determine $\mu_{a}$
directly, either confirming or rejecting the equivalence principle.

If $\mu_{a}$ exists for the neutron, it almost certainly exists for the other
neutral baryons. Whether it has an analogue for other neutral particles is not
as clear. For higher-spin mesons, such as the $\rho(770)^{0}$, with spin 1,
there may well be other ``anomalous acceleration moments.'' To see how such
moments would enter the Hamiltonian, it is necessary to examine the
acceleration generators for these particles.

\acknowledgments

The author wishes to thank Martin S. Altschul for his extensive and useful
comments.

\end{document}